\documentclass[10pt,letterpaper,twocolumn]{article}

\usepackage{ol2}
\usepackage{times}
\usepackage{hyperref}
\usepackage{graphicx}
\usepackage{longtable}
\usepackage{color}
\usepackage{amssymb}
\usepackage{srcltx}
\usepackage[latin1]{inputenc}
\usepackage[T1]{fontenc}

\usepackage{amsmath}

\newcommand{\eq}{\begin{equation}}
\newcommand{\ee}{\end{equation}}

\newcommand{\ket}[1]{| #1\rangle}

\begin{document}

\twocolumn[
\title{Deterministic qubit transfer between orbital and spin angular momentum of single photons}
\author{Vincenzo D'Ambrosio$^1$, Eleonora Nagali$^1$, Carlos H. Monken$^2$, Sergei Slussarenko$^3$, Lorenzo Marrucci$^3$, Fabio Sciarrino$^{1,4,*}$}
\address{$^1$Dipartimento di Fisica, Sapienza Universit\`{a} di Roma, Roma 00185, Italy

$^2$Departamento de Fisica, Universidade Federal de Minas Gerais, Caixa Postal 702, Belo Horizonte, MG 30161-970, Brazil

$^3$ Dipartimento di Scienze Fisiche, Universit\`{a} di
Napoli ``Federico II'', Compl.\ Univ.\ di Monte S. Angelo, 80126
Napoli, Italy

$^4$ Istituto Nazionale di Ottica Applicata, Firenze, Italy

$^*$Corresponding author: fabio.sciarrino@uniroma1.it}

\begin{abstract}
In this work we experimentally implement a deterministic transfer of
a generic qubit initially encoded in the orbital angular momentum of
a single photon to its polarization. Such transfer of quantum
information, completely reversible, has been implemented adopting a
electrically tunable q-plate device and a Sagnac interferomenter
with a Dove's prism. The adopted scheme exhibits a high fidelity and
low losses.
\end{abstract}

\ocis{(270.0270) Quantum Optics, (270.5585) Quantum information and processing }]

Qubits are often encoded in the polarization state of photons. This
is essentially due to the ease of manipulation and detection of the
spin angular momentum (SAM) of light. Besides SAM, photons can carry
orbital angular momentum (OAM) \cite{Alle92}, which is related to
the spatial distribution of the field. Photon states with a well
defined orbital angular momentum are the ones characterized by an
azimuthal dependence $e^{il\phi}$ of the phase front, where $l$ is
an integer (e.g. Laguerre-Gauss modes) \cite{Bazh92}. The Hilbert
space associated with the OAM degree of freedom is
infinite-dimensional, while the one associated with polarization is
restricted to two dimensions. This fact suggests the use of OAM,
alone or coupled with spin, as a resource to encode information in
higher dimensional quantum states, or qudits \cite{Moli07,
Naga10pra, Naga10prl}. A number of devices have been developed for
the generation and manipulation of OAM photon eigenstates, including
holograms \cite{Bazh92,Heck92}, mode converters \cite{Beij93},
spiral phase plates \cite{Beij94}, and more recently the
liquid-crystal q-plate (QP) \cite{Marr06}. The latter device, in
particular, introduces a controlled coupling between spin and
orbital angular momentum of a single photon, allowing for a coherent
transfer of information between the spaces associated with these two
degrees of freedom \cite{Naga09prl}. This feature has been recently
exploited for implementing a probabilistic quantum transferrer, i.e.
a device that can transfer a qubit from a degree of freedom to
another and \textit{vice versa} with a theoretical success
probability of 50$\%$ \cite{Naga09opt}. The transfer has been
demonstrated in particular from the bidimensional space of
polarization $\pi$ to a bidimensional subspace of OAM $o_{|l|}$. A
second qubit can then be added in the $\pi$ space, once that the
$\pi\rightarrow o_{|l|}$ transfer has been completed
\cite{Naga09prl,Naga10pra}. The probabilistic nature of the
demonstrated implementation is due to elements in the setup that
discard half of the information encoded in different OAM subspaces
($o_{|l|}\rightarrow\pi$) or in the polarization ($\pi\rightarrow o_{|l|}$).
However, a useful quantum information processing requires high
efficiencies. Therefore, the demonstration of a lossless
transferrer, ideally allowing for a qubit transfer with certainty
(success probability $p=1$), is an important goal. Schemes have been
proposed in order to achieve this goal \cite{Naga09opt}, but
hitherto they have not been demonstrated experimentally.

In this paper we report the experimental implementation of a
deterministic transferrer $o_2\rightarrow\pi$ based on a q-plate and
a polarizing Sagnac interferometer. In particular in this experiment
we employ q-plates with topological charge $q=1$ with tuning
controlled by an electric field, which allows to achieve a higher
efficiency of the device, and motorized wave-plates, so that the
transfer process is entirely automatized. Moreover, the same
experimental setup can be also used for the inverse, $\pi\rightarrow
o_2$ process, by reversing the propagation direction of light.

Let us first describe the working principle of the deterministic
transferrer, considering the $o_{|l|}\rightarrow\pi$ process. Let
us assume that the incoming photon is prepared in an arbitrary
OAM and fixed polarization state, so that all information is encoded in the OAM:
\eq
\ket{H}_{\pi}\ket{\phi}_{o_{|l|}}=\ket{H}_{\pi}\left(\alpha\ket{+l}+\beta\ket{-l}\right)_{o_{|l|}}
\label{pol} \ee 
where $H/V$ denotes the horizontal/vertical linear polarization. The state passes through 
a half waveplate (HWP) rotated at $\pi/8$ which transforms the polarization in a diagonal one:
$\ket{A}=\frac{\ket{H}+\ket{V}}{2}$, so that the state reads:
\begin{equation}
\ket{H}_{\pi}\left(\alpha\ket{+l}+\beta\ket{-l}\right)_{o_{|l|}}+\ket{V}_{\pi}\left(\alpha\ket{+l}+\beta\ket{-l}\right)_{o_{|l|}}
\label{input}
\end{equation}
Hereafter the
indices $\pi$ and $o_{|l|}$ are omitted for brevity. The
photon is then sent into a polarizing Sagnac interferometer (PSI) with a
polarizing beam-splitter (PBS) input/output port and a Dove prism
(DP) in one of its arms \cite{Naga09opt,Slus10}. Defining $\gamma$
as the angle between the base of the prism and the plane of the
interferometer, the action of the DP on the counter-propagating $H/V$ linear polarization components with
generic OAM $l$ is described by the following equations:
\begin{eqnarray}
\ket{H}\ket{l}&\rightarrow& e^{2il\gamma}\ket{H}\ket{l},\\
\ket{V}\ket{l}&\rightarrow& e^{-2il\gamma}\ket{V}\ket{l}
\end{eqnarray}
where the OAM-inverting effect of the reflections can be ignored,
for simplicity, as long as the total number of reflections in the
setup is even. Thus the two components of state \ref{input} in the PSI evolve as:
\begin{equation*}
\label{transfh}
    \alpha\ket{H}\ket{+l}+i\beta\ket{H}\ket{-l}\rightarrow
    \alpha e^{2i\gamma l}\ket{H}\ket{+l}+i\beta e^{-2i\gamma l}\ket{H}\ket{-l},
\end{equation*}
\begin{equation*}
\label{transfv}
\alpha\ket{V}\ket{+l}-i\beta\ket{V}\ket{-l}\rightarrow\alpha e^{-2i\gamma l}\ket{V}\ket{+l}-i\beta e^{2i\gamma l}\ket{V}\ket{-l}.
\end{equation*}
Setting $\gamma=\pi/(8 l)$ and applying these
transformations to state \ref{input}, one obtains (up to a global
phase factor) the output state:
\begin{equation*}
\alpha\ket{R}\ket{+l}+\beta\ket{L}\ket{-l}
\label{ent}
\end{equation*}
where $L/R$ denote left/right circular polarization. By passing through a q-plate, such state is hence transformed in:
\begin{equation*}
(\alpha\ket{L}+\beta\ket{R})\ket{0}_o=\ket{\phi}_{\pi}\ket{0}_o
\end{equation*}
that completes the transfer. Since all the intermediate
transformation steps are unitary, they are deterministic and
reversible. The inverse process $\pi\rightarrow o_{|l|}$, is therefore obtained
by simply inverting the light propagation through the same
components. It is also interesting to note that the action of the
transferrer is not limited to a $+l$ and $-l$ OAM subspace, but it
works with any pair $l_1 , l_2$ of OAM values. By repeating the
analysis above, one finds that the transfer is ensured as long as
the following general condition on the DP angle is satisfied: \eq
\gamma=\frac{\pi}{4(l_1-l_2)} \ee In this more general case,
however, the final polarization state is not $R$ but depends on the
values of OAM involved.

\begin{figure}[h]
\begin{center}
\includegraphics[scale=.33,bb=0 50 820 450, clip]{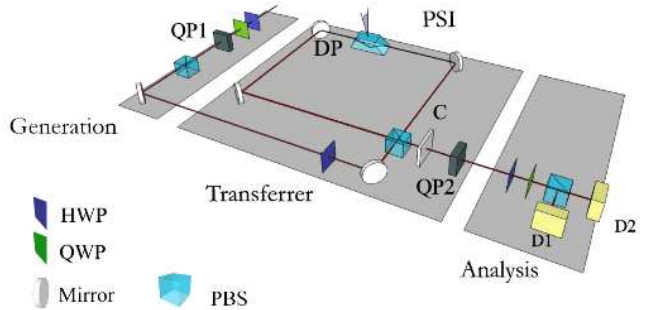}
\end{center}
\caption{Experimental setup adopted for the implementation of the
deterministic quantum transferrer $o_2\rightarrow\pi$. The input
photon, coming from the left, is prepared by a probabilistic
transferrer ($\pi\rightarrow o_2$) (first two waveplates, QP1 and
PBS) into an arbitrary $o_2$ state with polarization $H$. After this generation stage, the PSI and the
QP2 realize the deterministic transferrer ($o_2\rightarrow \pi$).
The outgoing polarization state is analyzed in the last part of the
setup (waveplates, PBS, detectors D1 and D2). C is a phase
compensation stage to correct all the unwanted phase shifts
introduced by the setup. All q-plates are electrically tuned.}
\label{setup}
\end{figure}

The experimental setup we used for demonstrating the deterministic
$o_2\rightarrow\pi$ transfer process can be divided in three
sections: (i) generation of single photons carrying the OAM input
qubit, (ii) quantum transferrer, and (iii) output state analysis
(see Fig.\ref{setup}).
In section (i) of the apparatus, an ultraviolet (UV) beam with
wavelength $\lambda_p=397.5$ nm pumps a $1.5$ mm thick nonlinear
crystal of $\beta$-barium borate (BBO), which generates, through the
spontaneous parametric down-conversion (SPDC) process, pairs of
photons in spatial modes $k_A$ and $k_B$ with the same wavelength
$\lambda=795$ nm and orthogonal linear polarizations $H$ and $V$.
The two photons in $k_A$ and $k_B$ are then spectrally purified by
interference filters with bandwidth $\Delta\lambda=3$ nm. The photon
in mode $k_A$ is detected and acts as a trigger of the single-photon
generation. The photon in mode $k_B$ is delivered to the main setup
via a single mode fiber, thus defining its transverse spatial mode
to a pure TEM$_{00}$, corresponding to OAM $l=0$. After the fiber
output, two wave plates compensate the polarization rotation
introduced by the fiber and a polarizing beam-splitter (PBS)
projects the photon onto the state $\ket{H}_{\pi}$. A quarter
waveplate (QWP) and a half-waveplate (HWP) are then used for
encoding an arbitrary qubit in the polarization degree of freedom of
the photon, as in Eq.\ (\ref{pol}). Finally, this
polarization-encoded qubit is converted into a OAM-encoded one using
the $\pi\rightarrow o_2$ \textit{probabilistic} transferrer, as
described in \cite{Naga09prl,Naga09opt}. For this step we used a
q-plate with $q=1$ combined with a PBS, providing conversion into
the photon state \eq \ket{\phi}_{o_2}=\alpha\ket{+2}+\beta\ket{-2}
\ee with a probability $p=0.5$. This completes the input state
preparation stage of our apparatus (generation box in
Fig.\ref{setup}).

This preparation stage was used in particular for generating all
states belonging to the three mutually unbiased bases spanning the
bidimensional OAM subspace with $l=\pm2$ ($o_2$):
$\{\ket{+2},\ket{-2}\},\{\ket{h},\ket{v}\},\{\ket{a},\ket{d}\}$,
where, analogously to the polarization case, we define linear
superpositions of $\ket{+2}$ and $\ket{-2}$ as
$\ket{h}=\frac{1}{\sqrt{2}}(\ket{+2}+\ket{-2}),
\ket{v}=\frac{1}{i\sqrt{2}}(\ket{+2}-\ket{-2})$ and
$\ket{a}=\frac{1}{\sqrt{2}}(\ket{h}+\ket{v}),
\ket{d}=\frac{1}{\sqrt{2}}(\ket{h}-\ket{v})$.

For the layout of the quantum transferrer setup, i.e. section (ii)
of the apparatus, we refer to Fig.\ \ref{setup}. The input qubit
$\ket{\phi}_{o_2}=\alpha\ket{+2}+\beta\ket{-2}$ prepared in the
previous section is now passed through a half wave plate in order to
set the polarization to the diagonal state
$\ket{A}$ and then injected in the PSI. The first PBS of the Sagnac
interferometer splits the two polarizations in two opposite
directions within the PSI, both passing through a DP rotated at angle
$\gamma=\frac{\pi}{16}$. The state is then sent through a q-plate
(QP2), which transforms the input state to
$\alpha\ket{L}+\beta\ket{R}=\ket{\varphi}_{\pi}$ with $l=0$ (mainly
TEM$_{00}$ mode). Thus the information initially encoded in the
orbital angular momentum has been transferred to the polarization
degree of freedom.

\begin{figure}[h]
\begin{center}
\includegraphics[scale=.3,bb=0 35 900 470, clip]{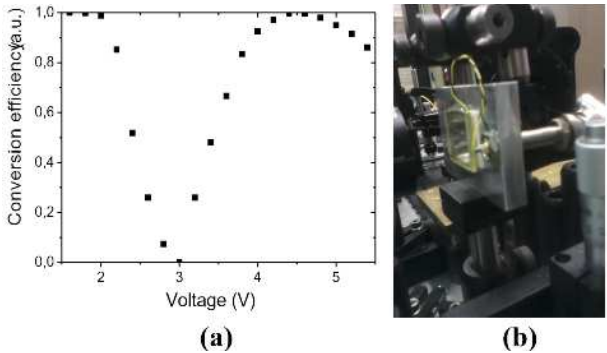}
\end{center}
\caption{\textbf{(a)} Conversion efficiency of the q-plate QP1 as a
function of the applied voltage. Above the threshold voltage (around
2.2 V), the conversion efficiency of the q-plate can be adjusted,
achieving its maximum around 4.5 V. \textbf{(b)} Photo of the
q-plate.} \label{qp}
\end{figure}

We note that both q-plates employed in this experiment are
electrically tunable. In this device, the q-plate birefringence
phase retardation $\delta$ is controlled by an external electric
field which changes the orientation of the liquid crystal molecular
director. This allows a more convenient control of the q-plate and a
faster time response as compared to the thermally-tuned q-plate
\cite{Kari09,Picc10}. By varying the applied voltage (with a sinusoidal signal) above the
Fr\'{e}edericksz threshold, the phase $\delta$ varies continuously
between 0 and $\pi$ (or more, depending on the cell thickness). This
in turn leads to a varying conversion efficiency of the q-plate,
related to $\delta$ by a sinusoidal relation \cite{Naga09opt}. When
$\delta=\pi$ the q-plate acts as a half wave plate on the
polarization, which corresponds to the maximum value of conversion
efficiency (tuned q-plate). In Fig.\ref{qp}-\textbf{a} we report as
an example the characteristic curve of QP1, where it is shown that
the optimal conversion efficiency is found to be around $4.5$ V.

After passing through the PSI and the second q-plate, all the
information encoded in the input qubit is transferred to the
polarization, and thus can be analyzed by a standard analysis setup
made of waveplates and a polarizing beam splitter, which form the
final section (iii) of our apparatus. The transmitted and reflected
photons from the PBS are coupled to single mode fibers and detected
by single photon counter modules $D_1$ and $D_2$. For full qubit
tomography, the output of the deterministic transferrer has been
analyzed in the three polarization bases
$\{\ket{R},\ket{L}\},\{\ket{H},\ket{V}\},\{\ket{A},\ket{D}\}$,
recording the coincidence counts between detectors $[D_1,D_T]$ and
$[D_2,D_T]$.
The overlap between the input OAM qubit and the polarization output
one after the transferrer has been estimated through the fidelity
parameter $F=\frac{C_{max}}{C_{max}+C_{min}}$. All results are
summarized in Table \ref{fid}, showing that very good values of
transfer fidelity are obtained.\\
\begin{table}[h!!]
\begin{small}
{\small
\begin{center}
\begin{tabular}{|c|c|}
\hline\hline
\textbf{State}& \textbf{Fidelity}  \\
\hline\hline
$\ket{+2}$ &  $(0.994\pm0.003)$\\
$\ket{-2}$ & $(0.992\pm0.003)$\\
$\ket{h}$ &  $(0.982\pm0.005)$\\
$\ket{v}$ &  $(0.944\pm0.008)$\\
$\ket{a}$ & $(0.992\pm0.003)$\\
$\ket{d}$ &  $(0.980\pm0.005)$\\
\hline
\textbf{Average value} &   $(0.980\pm0.002)$\\
\hline
\end{tabular}
\end{center}
} \caption{Experimental fidelity of the qubit transfer.} \label{fid}
\end{small}
\end{table}
Although
ideally the implemented setup has success probability $p=1$, the
actual value is limited by standard optical losses in the optical
components (mainly reflections, plus a little scattering and
absorption) and by the final single-mode fiber coupling step that we
used for experimental convenience and for mode purification, thus leading to
an overall efficiency of $0.324$. This value is three times larger than the
one achieved with the probabilistic device \cite{Naga09opt}. The obtained improvement 
is attributed to the adoption of the deterministic scheme based on the Sagnac interferometer,
to more efficient q-plates and to a better mode conversion exemplified by a higher single-mode 
coupling efficiency (compared to the one measured with previous q-plates) 
equal to $0.30$. As further improvements the reflection losses could be
reduced by adopting anti-reflection coating (in particular the q-plates are currently uncoated).
Finally we note the single-mode fiber coupling (currently $0.5$), although convenient for further 
processing of the output photons, is not a strictly required step.  

In summary, we have reported the experimental implementation of a
device that can transfer a qubit between the orbital angular
momentum and polarization degrees of freedom of single photons. The
ideal efficiency of the demonstrated device is one, so that the
device is theoretically deterministic. The scheme is based on the
combination of a q-plate with a Sagnac interferometer containing a
Dove prism. While the reported data refer to the OAM to spin qubit
transfer only, the same scheme can be used also to implement the
inverse process by simply reversing the direction of light
propagation in the same setup.

This work was supported by the Future and Emerging Technologies
(FET) programme within the Seventh Framework Programme for Research
of the European Commission, under FET-Open Grant No. 255914,
PHORBITECH. CHM acknowledges the financial support from CNPq
(Brazil).

\end{document}